\documentclass[twocolumn,showpacs,preprintnumbers,superscriptaddress,prl]{revtex4-1}

\usepackage{graphicx}

\usepackage{dcolumn}

\usepackage{bm}

\begin{document}

\preprint{}

\title{Disordered Ground State and Magnetic Field-Induced Long-Range Order in an $S$=3/2 Antiferromagnetic Honeycomb Lattice Compound Bi$_3$Mn$_4$O$_{12}$(NO$_3$)}

\author{M. Matsuda}

 \altaffiliation[Present address: ]{Neutron Scattering Science Division, Oak Ridge National Laboratory, Oak Ridge, Tennessee 37831, USA}
\affiliation{Quantum Beam Science Directorate, Japan Atomic Energy Agency (JAEA), Tokai, Ibaraki 319-1195, Japan}
 
\author{M. Azuma}

\affiliation{Institute for Chemical Research, Kyoto University, Uji, Kyoto 611-0011, 
Japan}

\author{M. Tokunaga}

\affiliation{Institute for Solid State Physics, University of Tokyo, Kashiwa, Chiba 277-8581, Japan}

\author{Y. Shimakawa}

\affiliation{Institute for Chemical Research, Kyoto University, Uji, Kyoto 611-0011, 
Japan}

\author{N. Kumada}

\affiliation{Department of Research Interdisciplinary Graduate School of Medicine and Engineering, University of Yamanashi, Miyamae-Cho 7, Kofu 400-8511, Japan}

\date{\today}

\begin{abstract}

Bi$_3$Mn$_4$O$_{12}$(NO$_3$), in which the Mn$^{4+}$ ions carry $S$=3/2, is the first honeycomb lattice system that shows no long-range magnetic order. Using neutron scattering, we have determined that short-range antiferromagnetic correlations develop at low temperatures. Applied magnetic fields induce a magnetic transition, in which the short-range order abruptly expands into a long-range order.

\end{abstract}

\pacs{75.25.-j, 75.30.Kz, 75.50.Ee}

\maketitle

Frustrated magnets show interesting phenomena originating from the macroscopic ground state degeneracy.~\cite{Diep} Exotic states can be chosen as the ground state from many possible candidates. Usually, geometrically frustrated systems consist of edge- and corner-shared triangles (triangular and kagom\'{e} lattices, respectively) and tetrahedra (pyrochlore lattice). However, even unfrustrated lattice systems, such as square or honeycomb lattice systems, can have frustrating interactions in the presence of antiferromagnetic further-neighbor interactions.

Magnetic field provides a perturbation to the degenerate ground states of frustrated magnets, sometimes giving rise to a novel phase transition.~\cite{ueda,rule,ross}  The Cr-based spinels, consisting of the pyrochlore lattice, show a half-magnetization plateau that is driven by the strong spin-lattice coupling.~\cite{ueda,penc}

The honeycomb lattice system with antiferromagnetic interactions has been previously studied. For example, Sr$L_2$O$_4$ ($L$=Gd, Dy, Ho, Er, Tm, and Yb) consists of distorted honeycomb latices.~\cite{karun} SrEr$_2$O$_4$ shows an anomalous long-range magnetic order in which only one of the two crystallographically inequivalent Er sites orders, probably due to the combined effects of exchange interactions, dipolar interactions, and crystal-field.~\cite{petrenko}

\begin{figure}
\includegraphics[width=8.5cm]{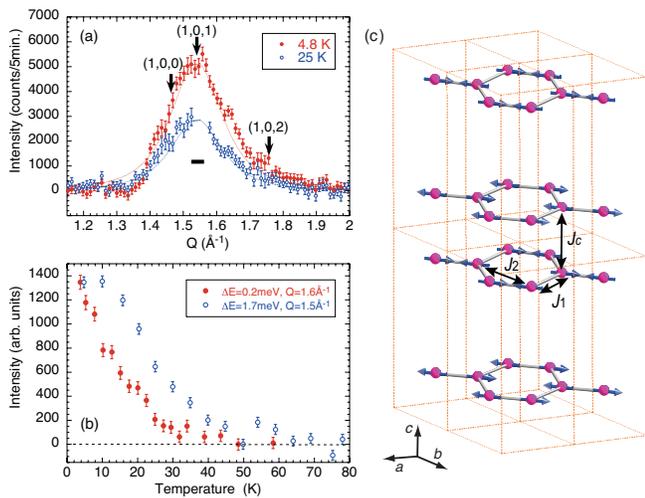}
\caption{(Color online) (a) Neutron powder diffraction patterns from Bi$_3$Mn$_4$O$_{12}$(NO$_3$) in zero magnetic field at 4.8 and 25 K. The broad peak originates from short-range antiferromagnetic correlations. The solid curves are the results of model calculations with $\xi_{ab}$=8 \AA\ and $\xi_c$=6 \AA. The thick horizontal bar represents the instrumental resolution. (b) The temperature dependence of the broad elastic magnetic signal measured with two different energy resolutions ($\Delta E$=0.2 and 1.7 meV). The two sets of the signals are normalized at the lowest temperature of each data. For all of the data shown here, the background data measured at $\sim$50 K was subtracted. (c) Magnetic structure in the magnetic field-induced phase. Mn$^{4+}$ ions in 2$\times$2$\times$2 unit cells are shown. The magnetic moments lie in the $ab$ plane, although the direction of the magnetic moments in the plane cannot be determined uniquely. Magnetic interactions $J_1$, $J_2$, and $J_c$ are also shown.}
\label{intensity}
\end{figure}

Bi$_3$Mn$_4$O$_{12}$(NO$_3$) has a trigonal structure ($P$3) with $a$=4.9692 \AA\ and $c$=13.1627 \AA\ and consists of regular honeycomb lattices of the magnetic Mn$^{4+}$ (nominally $t_{2g}^3; S = 3/2)$ ions.~\cite{azuma_JAC} Since a Jahn-Teller distortion is not observed in this compound, the orbital degrees of freedom appear to be unimportant. Therefore, this compound is a good model compound for an antiferromagnetic Heisenberg honeycomb lattice system. The magnetic susceptibility shows a broad maximum centered around 70 K, a characteristic feature of a two-dimensional antiferromagnet.
The Curie-Weiss temperature has a large value of $\Theta_{CW}$ = $-$257 K, indicating strong antiferromagnetic interactions. Long-range magnetic order is not observed down to 0.4 K. This is much lower than $|\Theta_{CW}|$ and indicates the presence of strong frustration \cite{azuma_JAC} and suggests that the nearest-neighbor ($J_1$) and next-nearest-neighbor interactions ($J_2$), as shown in Fig. \ref{intensity}, compete in this compound. As far as we know, Bi$_3$Mn$_4$O$_{12}$(NO$_3$) is the first honeycomb lattice system that shows no long-range order.

We performed neutron scattering experiments in a powder sample of Bi$_3$Mn$_4$O$_{12}$(NO$_3$). In zero magnetic field, it was confirmed that the compound does not show long-range magnetic order down to 3 K, indicating that the ground state is disordered. While no long range order is observed, short-range antiferromagnetic order, which is glassy in nature, develops at low temperatures.
It was also found that the coupling between the two adjacent honeycomb planes that form a bilayer structure ($J\rm_c$), as shown in Fig. \ref{intensity}, is not negligible, showing that Bi$_3$Mn$_4$O$_{12}$(NO$_3$) is a model compound for the bilayer honeycomb lattice system.
Under applied magnetic fields, this compound shows a commensurate long-range antiferromagnetic order, which corresponds to an expansion of the short-range order in zero field.
This indicates that magnetic field perturbs the degenerate ground state and gives rise to a N\'{e}el order.

A powder sample (4 g) of Bi$_3$Mn$_4$O$_{12}$(NO$_3$) was prepared by hydrothermal synthesis.~\cite{azuma_JAC} Although the sample contains 6.7 weight \% of MnO$_2$ as shown in Ref. \cite{azuma_JAC}, magnetic signals originating from the impurity phase were successfully removed.
Elastic neutron scattering experiments were carried out on the thermal and cold neutron triple-axis spectrometers TAS-2 and LTAS, respectively, installed at the guide hall of JRR-3 at Japan Atomic Energy Agency. The horizontal collimator sequences were guide-open-S-80'-open and guide-80'-S-80'-open on TAS-2 and LTAS, respectively. The fixed incident neutron energies were 14.7 and 3.5 meV with the energy resolutions ($\Delta E$) of 1.7 and 0.2 meV on TAS-2 and LTAS, respectively. Contamination from higher-order beams was effectively eliminated using PG and Be filters on TAS-2 and LTAS, respectively. A vertical magnetic field up to 10 T was applied using a new type of split-pair superconducting magnet that uses cryocoolers. For the experiments in magnetic fields, the powder sample was pressed into pellets to prevent the grains from orienting.
Inelastic neutron scattering experiments were carried out on TAS-2. The horizontal collimator sequence was guide-80'-S-80'-40'. The fixed final neutron energy was 14.7 meV with an energy resolution of $\sim$1.1 meV at $E$=0 meV.

Figure \ref{intensity}(a) shows neutron powder diffraction patterns for Bi$_3$Mn$_4$O$_{12}$(NO$_3$) at 4.8 and 25 K. A peak much broader than instrumental resolution ($\Delta Q\sim$0.04 \AA$^{-1}$) was observed at low temperatures. If the broad peak originates from a single honeycomb layer, an asymmetric peak is expected around (1, 0, 0) with a long tail at higher $Q$. This is not observed. Furthermore, there is no intense scattering at (1, 0, 2), indicating that the correlation between honeycomb bilayers is almost negligible. Since the observed peak position corresponds to (1, 0, 1), only one bilayer unit with antiferromagnetic arrangement in a honeycomb layer and also between two honeycomb layers contributes to the scattering. The broad peak was simulated using the powder averaged three-dimensional (3D) squared Lorentzian with an isotropic in-plane width, $A/[(Q-Q_{101})^2+2\xi_{ab}^{-2}+\xi_{c}^{-2}]^2$, where $A$, $\xi_{ab}$, and $\xi_{c}$ are constant, effective correlation lengths in the $ab$ plane and along the $c$ axis, respectively. The powder averaging was performed based on the method as described in Ref. \cite{warren}. Magnetic form factor was also included in the calculation. Since the peak width is much larger than the instrumental resolution, the resolution correction was not performed. The solid curves are the results of the calculations with $\xi_{ab}$=8 \AA\ and $\xi_c$=6 \AA. $\xi_{ab}$ is larger than one hexagon and $\xi_c$ is slightly longer than the distance between two honeycomb layers in one bilayer. The latter result is consistent with the above result that the (1, 0, 2) peak is almost absent. It is noted that the spectra at 4.5 and 20 K can be reproduced equally well by the same length parameters, indicating that the correlation length saturates at low temperatures.
\begin{figure}
\includegraphics[width=8.3cm]{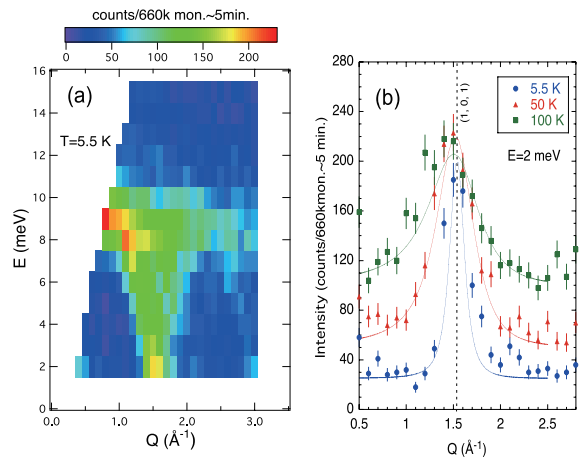}
\caption{(Color online) Inelastic neutron scattering spectra in zero magnetic field in Bi$_3$Mn$_4$O$_{12}$(NO$_3$) powder. (a) Image plot of the neutron scattering intensity in $E$-$Q$ space at 5.5 K. (b) Temperature dependence of the constant-energy scans measured at 2 meV. The solid curves are the results of the calculations with 3D squared Lorentzian.}
\label{ins}
\end{figure}

Figure \ref{intensity}(b) shows the temperature dependence of the broad magnetic signal around the peak position. The elastic magnetic signal increases almost monotonically below the characteristic temperature ($T\rm_f$), corresponding to a freezing temperature measured with the neutron energy scale. $T\rm_f$ strongly depends on the energy resolution of the neutrons. $T\rm_f$ becomes higher for measurements with broader energy resolution, with which signals from faster fluctuations are integrated. This is a typical behavior of the glassy magnetic systems in which spin fluctuations slow down gradually with lowering temperature. This is consistent with the results of susceptibility measurements that show a hysteresis between measurements with field-cool (FC) and zero-field-cool (ZFC) protocols.~\cite{azuma}

It was found that the ground state of Bi$_3$Mn$_4$O$_{12}$(NO$_3$) is disordered.
Our results indicate that $J_1$, $J_2$, and $J\rm_c$ are dominant
and there exist competing interactions between them. It is deduced from the short-range magnetic correlation that $J_1$ and $J\rm_c$ should be antiferromagnetic. In addition, $J_2$ should be antiferromagnetic and compete with $J_1$. In the honeycomb lattice with $J_1$ and $J_2$, the ground state is disordered when $\alpha$(=$J_2$/$J_1$)$>$1/6 and $\sim$0.15 for classical spins and $S$=3/2, respectively.~\cite{katsura,takano,kawamura}

Figure \ref{ins}(a) shows an image plot of inelastic neutron spectrum at $T$=5.5 K and $H$=0 T. Although there is no long-range magnetic order, relatively sharp magnetic excitations were observed. The spin-wave-like excitation rises from $Q\sim$1.5 \AA$^{-1}$, corresponding to the (1, 0, 1) position, and the band width of the excitation is $\sim$9 meV. An excitation gap originating from the magnetic anisotropy was not observed even with high energy resolution experiments with $\Delta E\sim$0.2 meV, indicating that the interactions are isotropic.

\begin{figure}
\includegraphics[width=8.5cm]{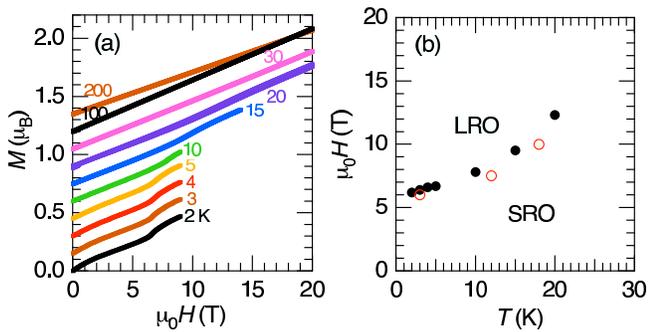}
\caption{(Color online) (a) Magnetization versus magnetic field in dc ($T\le$10 K) and pulsed  fields ($T\ge$15 K) for Bi$_3$Mn$_4$O$_{12}$(NO$_3$) in a temperature range between 2 and 200 K. (b) Temperature-magnetic field phase diagram. Filled and open circles show the data points determined from the magnetization and neutron scattering measurements, respectively. LRO and SRO stand for the long-range and short-range ordered phases, respectively.}
\label{mh_curve}
\end{figure}
\begin{figure}
\includegraphics[width=5.4cm]{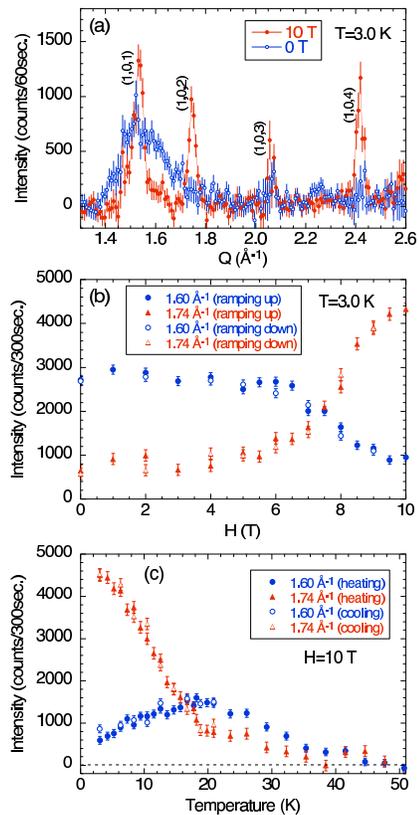}
\caption{(Color online) (a) Neutron powder diffraction patterns in Bi$_3$Mn$_4$O$_{12}$(NO$_3$) with $\Delta E\sim$1.7 meV at $H$=0 and 10 T and at $T$=3 K. At 10 T the broad magnetic peak is reduced and sharp magnetic peaks appear. The magnetic field (b) and temperature dependences (c) of the elastic magnetic signal at $Q$=1.60 and 1.74 \AA$^{-1}$. The signals at 1.60 and 1.74 \AA$^{-1}$ mostly originate from the short-range order and the 3D long-range order, respectively. In all the data shown here, the background signal was subtracted.}
\label{field_dep}
\end{figure}
In order to clarify the temperature dependence of the spin fluctuations in more detail, temperature dependence of the constant-energy scans was measured at 2 meV as shown in Fig. \ref{ins}(b).
The peak width becomes broader with increasing temperature.
The constant background intensity, which probably originates from paramagnetic scattering and a tail of nuclear incoherent scattering, increases with increasing temperature. The inverse peak width corresponds to the correlation length. The solid curve at 5.5 K is the result of calculations with $\xi_{ab}$=8 \AA\ and $\xi_c$=6 \AA\ as in the elastic data shown in Fig. \ref{intensity}(a). The inelastic peak is broader than that of the elastic data, probably because the excitations are dispersive and also because the $Q$ resolution is broader at 2 meV.
Since the peak profile is symmetric at higher temperatures, indicative of isotropic spin correlations, the data at 50 and 100 K were simulated using the powder averaged isotropic 3D squared Lorentzian.
The solid curves are the results of the calculations with $\xi$=3.3 and 2.5 \AA\ at 50 and 100 K, respectively. Even though the correlation length is very short, the peak is almost symmetric, suggesting that $J\rm_c$ is as large as $J_1$.

From our inelastic neutron scattering experiments, the interactions can be estimated roughly. The maximum of the magnetic excitation, $\sim$9 meV, corresponds to the sum of all the interactions multiplied by number of bonds and $S$(=3/2), $S(3J_1+6J_2+J\rm_c)$, indicating that $3J_1+6J_2+J\rm_c\sim$6 meV. Suppose that $J_2$/$J_1$=0.15 corresponding to the critical value and $J\rm_c$=0.5$J_1$, it would be estimated that $J_1$=2$J\rm_c$=1.4 meV and $J_2$=0.20 meV. For comparison, $J_1$ is estimated to be 3.0 meV from $\Theta_{CW}$ (= $-$257 K), when $J_2$ and $J\rm_c$ are not considered. The agreement is reasonably good, since $J_1$ is reduced when $J_2$ and $J\rm_c$ are included.

Another characteristic feature in Bi$_3$Mn$_4$O$_{12}$(NO$_3$) is that magnetization shows a small jump of $\sim$0.1$\mu\rm_B$/f.u. around 6-13 T below 20 K, as shown in Fig. \ref{mh_curve}(a), indicating a transition to a metamagnetic phase.
In order to clarify the nature of the magnetic field-induced phase, we performed neutron scattering experiments under applied magnetic fields. 

Figure \ref{field_dep}(a) shows the neutron powder diffraction patterns in Bi$_3$Mn$_4$O$_{12}$(NO$_3$) at $H$=0 and 10 T. The spectrum at 0 T shows a broad peak originating from the short-range antiferromagnetic order as described above.
With application of magnetic field at $T$=3 K, the broad magnetic intensity is reduced and resolution-limited sharp magnetic Bragg peaks appear above $H\sim$6 T, as shown in Fig. \ref{field_dep}(b). This result indicates that long-range magnetic order is induced in magnetic field.
Figure \ref{field_dep}(c) shows the temperature dependence of the elastic magnetic signals at $Q$=1.60 and 1.74 \AA$^{-1}$ at $H$=10 T. The signal originating from the long-range order vanishes and that from the short-range order recovers around 18 K.

Since the most intense peak is the (1, 0, 1) reflection, the overall spin correlations do not change even in the magnetic field-induced phase. In the $ab$ plane, the nearest-neighbor spins align antiparallel. The adjacent spins along the $c$ axis also align antiparallel. Therefore, there is no frustration in the magnetic structure. Since the width of the magnetic field-induced sharp peak is almost the same as the instrumental resolution, it is evident that the magnetic structure is long-ranged and commensurate. From the results of the magnetic structural analysis, we determined that the spins lie in the $ab$ plane and that the ordered moment is $\sim$1.8$\mu\rm_B$ at 3 K. The magnetic structure is shown in Fig. \ref{intensity}(c). The moment size is about 2/3 of the full moment 3$\mu\rm_B$ for the Mn$^{4+}$ ion as reported in Ref. \cite{azuma_JAC}. As shown in Fig. \ref{field_dep}(a), there still exists a broad signal around 1.6 \AA$^{-1}$ at 10 T. This is the remanent of the broad peak observed in the zero magnetic field. This indicates that the field-induced transition depends on magnetic field direction and that only the magnetic moments perpendicular to the magnetic field contribute to the ordering and the others remain disordered. It is puzzling why the transition depends on the field direction, although the compound is an isotropic Heisenberg system. The Dyaloshinski-Moriya (DM) interaction may be an origin of the anisotropy, since the crystal structure ($P$3) does not have inversion symmetry.

The magnetization measurements show that the metamagnetic phase has a ferromagnetic component of $\sim$0.1$\mu\rm_B$/f.u. A canting of spins probably occurs in the field-induced phase, although the small canted component cannot be detected in the present neutron diffraction measurements. It is possible that the canting is driven by the DM interaction, as described above. The stabilization of the magnetic ordering due to the spin canting in the magnetic moment is probably the origin of the field induced long-range ordering.

The magnetic field-temperature phase diagram are summarized in Fig. \ref{mh_curve}(b). We observed the magnetic field-induced phase transition, in which the short-range order expands into the long-range N\'{e}el order by applying magnetic field.
This indicates that this compound is located near the boundary of the ordered and disordered phases with $\alpha$ close to 0.15. It was reported in pyroclore systems Tb$_2$Ti$_2$O$_7$ and Yb$_2$Ti$_2$O$_7$ that magnetic field along a particular direction induces a magnetic transition to a long-range ordered phase although the detailed structure is unknown.~\cite{rule,ross} The transition temperature increases with increasing magnetic field in these compounds. As shown in Fig. \ref{mh_curve}(b), the same behavior was observed in Bi$_3$Mn$_4$O$_{12}$(NO$_3$). This indicates that the collinear N\'{e}el state becomes more stabilized at higher temperatures. This may suggest that thermal fluctuations stabilize the N\'{e}el order, where "order by disorder" mechanism is at work.

We have observed disordered ground state and the magnetic field-induced transition to commensurate N\'{e}el order in the bilayer honeycomb system Bi$_3$Mn$_4$O$_{12}$(NO$_3$).
We emphasize that the zero-field and field-induced spin correlations in the honeycomb lattice system were studied experimentally for the first time.

We would like to thank Y. Motome, S. Okumura, and H. Kawamura for stimulating discussions. This work was partially supported by Grant-in-Aid for Scientific Research on Priority Areas "Novel States of Matter Induced by Frustration" (19052008) and "High Field Spin Science in 100T" (No.451) from the Ministry of Education, Culture, Sports, Science and Technology (MEXT) of Japan.

\end{document}